\documentclass[twocolumn,preprintnumbers,amsmath,aps]{revtex4}
\usepackage{graphicx}
\usepackage{dcolumn}
\usepackage{bm}
\usepackage{subfigure}
\usepackage{color}
\begin{document}
\newcommand{\kvec}{\mbox{{\scriptsize {\bf k}}}}
\def\eq#1{Eq.\hspace{1mm}(\ref{#1})}
\def\fig#1{Fig.\hspace{1mm}\ref{#1}}
\def\tab#1{Tab.\hspace{1mm}\ref{#1}}

\title{
---------------------------------------------------------------------------------------------------------------\\
The correlation between the energy gap and the pseudogap temperature in cuprates: 
the YCBCZO and LSHCO case}
\author{R. Szcz{\c{e}}{\'s}niak, M.W. Jarosik, A.M. Duda}
\affiliation{Institute of Physics, Cz{\c{e}}stochowa University of Technology, Al. Armii Krajowej 19, 42-200 Cz{\c{e}}stochowa, Poland}
\email{jarosikmw@wip.pcz.pl}
\date{\today} 
\begin{abstract}
The paper analyzes the influence of the hole density, the out-of-plane or in-plane disorder, and the isotopic oxygen mass on the zero temperature energy gap ($2\Delta\left(0\right)$) for $\rm{Y}_{1-x}\rm{Ca}_{x}\rm{Ba}_2\left(\rm{Cu}_{1-y}\rm{Zn}_{y}\right)_{3}\rm{O}_{7-\delta}$ (YCBCZO) and $\rm{La}_{1.96-x}\rm{Sr}_{x}\rm{Ho}_{0.04}\rm{CuO}_{4}$ (LSHCO) superconductors. It has been found that the energy gap is visibly correlated with the value of the pseudogap temperature ($T^{\star}$). On the other hand, no correlation between $2\Delta\left(0\right)$ and the  critical temperature ($T_{C}$) has been found. The above results mean that the value of the dimensionless ratio $2\Delta\left(0\right)/k_{B}T_{C}$ can vary very strongly together with the chemical composition, while the parameter $2\Delta\left(0\right)/k_{B}T^{\star}$ does not change significantly. In the paper, the analytical formula which binds the zero temperature energy gap and the pseudogap temperature has been also presented.
\end{abstract}
\maketitle

\vspace*{0.25 cm} 

The superconductivity in the compounds of copper oxides (cuprates) was discovered in 1986 by Bednorz and Muller \cite{Bednorz}. It is now known that in the family of cuprates  the compounds of the highest critical temperatures ($T_{C}$) exist. For example, in the ${\rm HgBa_{2}Ca_{2}Cu_{3}O_{8+\delta}}$  (HBCCO) superconductor under the pressure at $31$ GPa, the critical temperature equals about $164$ K \cite{Gao}. However, Takeshita {\it et al.} have reported recently that the correct maximum value of the critical temperature for HBCCO is a little bit lower, and it appears at much lower pressure ($T_{C}=153$ K at $15$ GPa) \cite{Takeshita}.

The thermodynamics of the high-temperature superconducting state in cuprates differs significantly from the thermodynamics predicted in the framework of the classical BCS theory \cite{Bardeen}, \cite{Bardeen01}. In addition to the too high value of the critical temperature, the most important difference seems to be in the existence of the second characteristic temperature, which is called the pseudogap temperature ($T^{\star}$). 

Currently, it is believed that the critical temperature in cuprates sets the maximum value of $T$, at which disappears the coherence of the superconducting state, while $T^{\star}$ determines the temperature, in which the energy gap ($2\Delta$) ceases to exist at the Fermi level \cite{Kanigel}. 

It should be noted that both temperatures are equal in the classical BCS theory, wherein the theory predicts the universal relationship between the value of the zero temperature energy gap and the critical temperature: $2\Delta\left(0\right)/k_{B}T_{C}=3.53$, where $k_{B}$ is the Boltzmann constant \cite{Carbotte}.

In the presented paper, we have examined the impact of the various factors (the hole density ($p$), the disorder, and the oxygen isotopic mass) on the energy gap in the cuprates $\rm{Y}_{1-x}\rm{Ca}_{x}\rm{Ba}_2\left(\rm{Cu}_{1-y}\rm{Zn}_{y}\right)_{3}\rm{O}_{7-\delta}$ (YCBCZO) and  $\rm{La}_{1.96-x}\rm{Sr}_{x}\rm{Ho}_{0.04}\rm{CuO}_{4}$ (LSHCO) \cite{Naqib}, \cite{Naqib01}, \cite{Hafliger}. 

The primary objective of the study was to determine the relationship between the zero temperature energy gap and $T_{C}$ or $T^{\star}$. 
The obtained results allowed then the determination of the values of the dimensionless ratios:  $R_{\Delta}\equiv 2\Delta(0)/k_{B}T_{C}$ and 
$R_{\Delta^{\star}}\equiv 2\Delta(0)/k_{B}T^{\star}$. 

In the last step, we have derived the analytical formula which binds the energy gap and the pseudogap temperature.   

\vspace*{0.25cm}

All calculations have been performed in the framework of the theory, which assumes that the pairing mechanism in cuprates is induced by the 
electron-phonon interaction and the electron-electron correlations renormalized by the phonons. Additionally, through the appropriate selection of the electron band energy, the influence of the quasi-two-dimensionality of the electron system (the cooper-oxygen plane) on the physical properties of the studied compounds has been taken into account. 

The detailed description of the considered theory, together with the corresponding analysis leading to the fundamental thermodynamic equation, the reader may find in the paper \cite{Radek01}. Additional information is also contained in the following works: \cite{Radek02} (analysis of the ARPES method), \cite{Radek03} (thermodynamics and ARPES for $({\rm Hg}_{1-x}{\rm Sn}_{x}){\rm Ba}_{2}{\rm Ca}_{2}{\rm Cu}_{3}{\rm O}_{8+\delta}$), and \cite{Radek04} 
(thermodynamics of the high-temperature superconductors with the maximum $T_{C}$).


\vspace*{0.25cm}

In the cases considered in the presented paper, the fundamental thermodynamic equation that determines the properties of the high-temperature superconducting state of $d$-wave symmetry has the following form \cite{Radek01}:
\begin{equation}
\label{r1}
1=\left(V^{\left(\eta\right)}+\frac{U^{\left(\eta\right)}}{6}|\overline\Delta^{\left(\eta\right)}|^{2}\right)\frac{1}{N_{0}}\sum^{\omega_{0}}_{\kvec}
\frac{\eta^{2}(\bf k)}{2E_{\kvec}^{\left(\eta\right)}}\tanh\frac{\beta E_{\kvec}^{\left(\eta\right)}}{2},
\end{equation}
where the pairing potentials for the electron-phonon and electron-electron-phonon interaction have been denoted as 
$V^{\left(\eta\right)}$ and $U^{\left(\eta\right)}$, respectively. 
The quantity $\overline\Delta^{\left(\eta\right)}$ is the amplitude of the order parameter for $d$-wave symmetry:  $\eta{\left(\bf k\right)}\equiv 2\left[\cos\left(k_{x}\right)-\cos\left(k_{y}\right)\right]$. 

The symbol $E_{\kvec}^{\left(\eta\right)}$ is defined by the expression below: 
\begin{equation}
\label{r2}
E_{\kvec}^{\left(\eta\right)}\equiv\sqrt{\varepsilon_{\kvec}^{2}+\left(V^{\left(\eta\right)}+ \frac{U^{\left(\eta\right)}}{6}|\overline\Delta^{\left(\eta\right)}|^{2}\right)^{2}\left(|\overline\Delta^{\left(\eta\right)}|\eta\left(\bf k\right)\right)^{2}}, 
\end{equation}
where the function $\varepsilon_{\kvec}$ determines the electron band energy: $\varepsilon_{\kvec}=-t\gamma\left(\bf k\right)$; $t$ denotes the hopping integral and $\gamma\left({\bf k}\right)\equiv 2\left [\cos\left(k_{x}\right)+\cos\left(k_{y}\right)\right]$. 

The inverted temperature ($\beta$) is given by the expression: $\beta\equiv 1/k_{B}T$. 

The normalization constant is given by: $N_{0}\equiv 1/\sum^{\omega_{0}}_{\kvec}$.
The symbol $\omega_{0}$ represents the characteristic phonon frequency, which is of the order of Debye frequency.

Note that the sum over the momentums in \eq{r1} should be replaced with the integral in the following manner:   $\sum^{\omega_{0}}_{\kvec}\simeq
\int^{\pi}_{-\pi}\int^{\pi}_{-\pi}dk_{x}dk_{y}\theta\left(\omega_{0}-\left|\varepsilon_{\left(\kvec_{x},\kvec_{y}\right)}\right|\right)$, where $\theta$ is the Heaviside function.

In order to simplify the numerical calculations and perform the analytical calculations in the subsequent part of the work, \eq{r1} should be converted into a more convenient form. For this reason, we have introduced the designations: 
$v^{2}\equiv V^{\left(\eta\right)}$,
$u^{2}\equiv \frac{U^{\left(\eta\right)}}{6}$ and:
\begin{equation}
\label{r3}
\Delta\equiv \left(v^{2}+u^{2}|\overline\Delta^{\left(\eta\right)}|^{2}\right)|\overline\Delta^{\left(\eta\right)}|.
\end{equation}

Now, \eq{r1} can be rewritten in the following way: 
\begin{equation}
\label{r4}
1=\left(v^{2}+u^{2}|\overline\Delta^{\left(\eta\right)}|^{2}\right)I^{\left(\eta\right)}\left(\Delta,T\right),
\end{equation}
where:
\begin{eqnarray}
\label{r5}
I^{\left(\eta\right)}\left(\Delta,T\right)&\equiv& 
\frac{1}{N_{0}}\sum^{\omega_{0}}_{\kvec}
\frac{\eta^{2}(\bf k)}{2\sqrt{\varepsilon_{\kvec}^{2}+\left(\eta\left(\bf k\right)\Delta\right)^{2}}}\\ \nonumber
&\times&
\tanh\frac{\beta\sqrt{\varepsilon_{\kvec}^{2}+\left(\eta\left(\bf k\right)\Delta\right)^{2}}}{2}.
\end{eqnarray}

Using the expression (3), we can make the following transformation of \eq{r4}:
\begin{equation}
\label{r6}
1=\left[v^{2}+\left[\frac{u\Delta}{v^{2}+\left[\frac{u\Delta}{v^{2}+\left[...\right]^{2}}\right]^{2}}\right]^{2}\right]
I^{\left(\eta\right)}\left(\Delta,T\right).
\end{equation}
\begin{figure}
\includegraphics[width=\columnwidth]{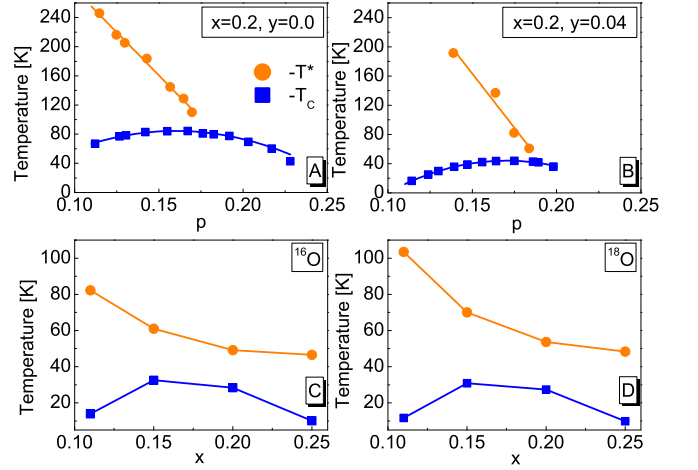}
\caption{(A)-(B) The critical temperature and the pseudogap temperature as a function of the hole density for YCBCZO superconductor \cite{Naqib}, \cite{Naqib01}. The results have been plotted for the systems with different degrees of disorder. 
(C)-(D) The dependence of the critical temperature and the pseudogap temperature on the concentration of strontium for LSHCO superconductor 
(the out-of-plane disorder). Additionally, two isotopes of oxygen have been taken into account \cite{Hafliger}.}
\label{f1}
\end{figure}

It turns out that \eq{r6} can be written in the compact form: 

\begin{equation}
\label{r7}
1=\left[v^{2}+\left(u\Delta\right)^{2}\left(I^{\left(\eta\right)}\left(\Delta,T\right)\right)^{2}\right]
I^{\left(\eta\right)}\left(\Delta,T\right).
\end{equation}

The equivalence of equations (6) and (7) can be most easily proven when determining the quantity $I^{\left(\eta\right)}\left(\Delta,T\right)$ 
from \eq{r7} and then reinserting the resulting formula in the square brackets in \eq{r7}.

\vspace*{0.25cm}

The input parameters for \eq{r7} are as follows: the hopping integral, the characteristic phonon frequency and the pairing potentials. 

The same values of $t$ and $\omega_{0}$ for the YCBCZO superconductor have been assumed as for the compound    
${\rm YBa_{2}Cu_{3}O_{7-\delta}}$ (YBCO): $t=250$ meV and $\omega_{0}=75$ meV \cite{Nunner}, \cite{Bohnen}. In the case of the LSHCO superconductor, we have 
based on the values of $t$ and $\omega_{0}$ obtained for ${\rm La_{2-x}Sr_{x}CuO_{4}}$ (LSCO): $t=240$ meV and $\omega_{0}=96$ meV \cite{Xu}, \cite{Kim}.

The pairing potentials $v$ and $u$ have been chosen in such a way that the values of the critical temperature and the pseudogap temperature calculated on the basis of \eq{r7} would agree with the experimental values of $T_{C}$ and $T^{\star}$ determined in the works \cite{Naqib}, \cite{Naqib01} and \cite{Hafliger}. It should be noted that this can be done in a relatively simple way, because the electron-phonon potential is the unique function of the critical temperature ($v=v\left(T_{C}\right)$). Then, we have been able to determine the renormalized potential of the electron-electron interaction: $u=u\left(v\left(T_{C}\right),T^{\star}\right)$ \cite{Radek01}. The values of $T_{C}$, $T^{\star}$ and the corresponding results have been presented in \fig{f1} and in \tab{t1}.

\begin{table}[t]
\renewcommand*{\tablename}{\small{Tab.}}
\renewcommand*{\baselinestretch}{1}
\caption{\label{t1} The critical temperature, the pseudogap temperature, and the values of the pairing potentials $v$ and $u$ for the YCBCZO and LSHCO superconductors. In the case of YCBCZO, the hole density ($p$) has been calculated based on the expression:  
$T_{C}\left(p\right)/T_{C,{\rm max}}=1-p_{A}\left(p-p_{B}\right)^{2}$ \cite{Presland}. Let us note that for the superconductors, which do not show the in-plane disorder, it has been obtained: $p_{A}=82.6$ and $p_{B}=0.16$. In the opposite case, the values of the parameters $p_{A}$ and $p_{B}$ increase together with the increasing disorder \cite{Naqib03}.}
\begin{center}
\begin{tabular}{|c|c|c|c|c|c|}\hline
                          &           &      &       &      &        \\
{\bf Material} & 
 ${\bf Type}$ & 
${\bf T_{C}}$&
${\bf T^{\star}}$ &
${\bf v}$ &
${\bf u }$   
\\
 &           &   ${\bf\left[K\right]}$   &   ${\bf\left[K\right]}$   &   ${\bf\left[\sqrt{meV}\right]}$   &     ${\bf\left[\sqrt{meV}\right]}$   \\
                         &           &      &       &      &        \\
\hline
                          &           &      &       &      &        \\
YCBCZO    \\ (x=0.2, y=0) & $p$=0.115 & 70.1 & 243.9 & 2.2371 & 2.9926  \\ 
                          & $p$=0.123 & 74.9 & 224.4 & 2.2749 & 2.8642  \\ 
                          & $p$=0.131 & 78.5 & 205.5 & 2.3005 & 2.7422  \\
                          & $p$=0.139 & 81.3 & 186.7 & 2.3215 & 2.6133  \\
                          & $p$=0.147 & 83.2 & 167.8 & 2.3354 & 2.4868  \\
                          & $p$=0.155 & 84.2 & 148.9 & 2.3433 & 2.3548  \\
                          & $p$=0.163 & 84.3 & 130.0 & 2.3442 & 2.2112  \\
                          & $p$=0.170 & 83.7 & 114.2 & 2.3393 & 2.0922  \\ 
\hline
                            &           &       &       &      &       \\
YCBCZO    \\ (x=0.2, y=0.04)& $p$=0.139 & 35.4  & 196.5 & 1.9159 & 2.9381 \\ 
                            & $p$=0.146 & 38.9  & 175.2 & 1.9503 & 2.8135 \\
                            & $p$=0.153 & 41.5  & 154.4 & 1.9811 & 2.6883 \\ 
                            & $p$=0.160 & 43.2  & 133.5 & 2.0009 & 2.5708 \\ 
                            & $p$=0.167 & 44.0  & 112.7 & 2.0101 & 2.4508 \\
                            & $p$=0.174 & 43.9  & 91.9  & 2.0097 & 2.3550 \\
                            & $p$=0.181 & 43.0  & 71.1  & 1.9993 & 2.2204 \\
                            & $p$=0.184 & 42.4  & 62.9  & 1.9929 & 2.1765 \\
\hline
                            &        &      &      &         &       \\
LSHCO    \\ ($^{16}\rm{O}$) & x=0.11 & 13.9 & 82.2 & 1.7435  & 3.0676 \\
                            & x=0.15 & 32.5 & 61.0 & 2.0325  & 2.7586 \\
                            & x=0.20 & 28.4 & 49.2 & 1.9836  & 2.7804 \\
                            & x=0.25 & 10.1 & 46.5 & 1.6474  & 3.0757 \\ 
\hline
                            &        &      &       &         &       \\
LSHCO    \\ ($^{18}\rm{O}$) & x=0.11 & 11.7 & 103.5 & 1.6924  & 3.1509 \\
                            & x=0.15 & 30.8 & 70.0  & 2.0100  & 2.8137 \\
                            & x=0.20 & 27.3 & 53.7  & 1.9700  & 2.8069 \\
                            & x=0.25 & 9.8  & 48.4  & 1.6331  & 3.0867 \\ 
\hline
\end{tabular}
\end{center}
\end{table}

On the basis of \fig{f1} (A) and the results collected in the work \cite{Radek04}, it can be easily seen that the out-of-plane disorder induced in YBCO by calcium does not change significantly $T_{C}$ and $T^{\star}$. For this reason, the values of the potentials $v$ and $u$ obtained for the YCBCZO compound (x=0.2, y=0) are very close to the values of the pairing potentials received for the ordered YBCO compound. 

\begin{figure}[ht]
\includegraphics[width=\columnwidth]{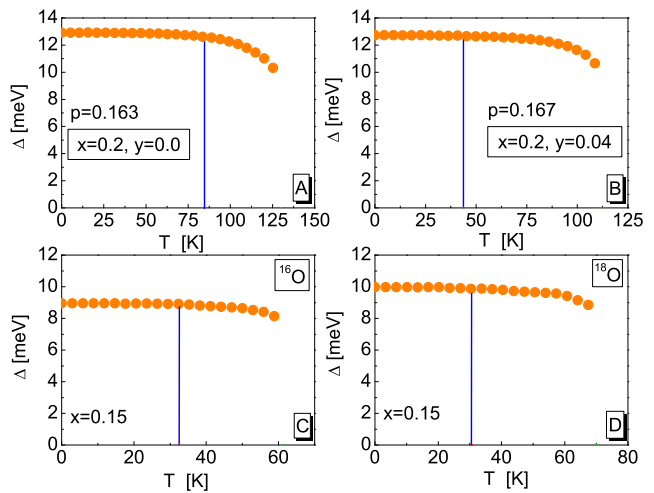}
\caption{(A)-(B) The order parameter as a function of the temperature for the YCBCZO superconductor. The selected values of doping have been adopted.  
         (C)-(D) The temperature dependence of the order parameter for the LSHCO superconductor. The blue vertical line in the drawings sets the value of the critical temperature. 
        }
\label{f2}
\end{figure}
\begin{figure}[ht]
\includegraphics[width=\columnwidth]{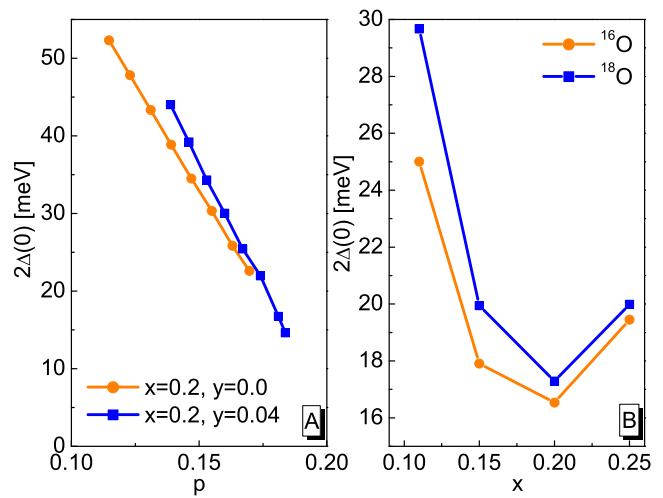}
\caption{(A) The zero temperature energy gap as a function of the hole density for the YCBCZO superconductor. 
(B) The value of the zero temperature energy gap in the dependence on the concentration of strontium for the LSHCO superconductor.}
\label{f3}
\end{figure}

In the case of the in-plane disorder, generated by zinc in YCBCZO (x=0.2, y=0.04), the characteristic temperatures $T_{C}$ and $T^{\star}$ strongly decrease together with the increasing concentration of Zn  (\fig{f1} (B)). As a result, the value of the pairing potential for the electron-phonon interaction significantly decreases. On the other hand, $u$ changes only slightly, because it must compensate for the strong decline of $v$, so the experimental value $T^{\star}$ can be reproduced. 

\begin{table}[t]
\renewcommand*{\tablename}{\small{Tab.}}
\renewcommand*{\baselinestretch}{1}
\caption{\label{t2} The values of the zero temperature energy gap at the Fermi level, the ratio $R_{\Delta}$, and the ratio $R_{\Delta^{\star}}$ for the 
YCBCZO and LSHCO superconductors.}
\begin{center}
\begin{tabular}{|c|c|c|c|c|}\hline
                          &           &        &        &     \\
{\bf Material} & 
 ${\bf Type}$ & 
${\bf 2\Delta\left(0\right)}$ &
${\bf R_{\Delta}}$ &
${\bf R_{\Delta^{\star}}}$     
\\
                          &           &    ${\bf \left[meV\right]}$    &        &     \\
                          &           &        &        &     \\
\hline
                          &           &        &        &     \\
YCBCZO    \\ (x=0.2, y=0) & $p$=0.115 & 52.30  & 8.66   & 2.49\\ 
                          & $p$=0.123 & 47.83  & 7.41   & 2.47\\ 
                          & $p$=0.131 & 43.35  & 6.41   & 2.45\\
                          & $p$=0.139 & 38.86  & 5.55   & 2.42\\
                          & $p$=0.147 & 34.49  & 4.81   & 2.39\\
                          & $p$=0.155 & 30.34  & 4.18   & 2.36\\
                          & $p$=0.163 & 25.84  & 3.56   & 2.31\\
                          & $p$=0.170 & 22.60  & 3.13   & 2.30\\ 
\hline
                            &           &        &       & \\
YCBCZO    \\ (x=0.2, y=0.04)& $p$=0.139 & 44.04  & 14.43 & 2.60\\ 
                            & $p$=0.146 & 39.21  & 11.69 & 2.60\\
                            & $p$=0.153 & 34.32  & 9.60  & 2.58\\ 
                            & $p$=0.160 & 30.04  & 8.08  & 2.61\\ 
                            & $p$=0.167 & 25.46  & 6.72  & 2.62\\
                            & $p$=0.174 & 21.98  & 5.80  & 2.77\\
                            & $p$=0.181 & 16.73  & 4.51  & 2.73\\
                            & $p$=0.184 & 14.63  & 4.00  & 2.70\\
\hline
                            &        &       &         & \\
LSHCO    \\ ($^{16}\rm{O}$) & x=0.11 & 25.00 & 20.87   & 3.53\\
                            & x=0.15 & 17.90 & 6.39    & 3.41\\
                            & x=0.20 & 16.54 & 6.76    & 3.90\\
                            & x=0.25 & 19.45 & 22.35   & 4.85\\ 
\hline
                            &        &       &       & \\
LSHCO    \\ ($^{18}\rm{O}$) & x=0.11 & 29.67 & 29.43 & 3.33\\
                            & x=0.15 & 19.94 & 7.51  & 3.31\\
                            & x=0.20 & 17.28 & 7.35  & 3.74\\
                            & x=0.25 & 19.99 & 23.67 & 4.79\\ 
\hline
\end{tabular}
\end{center}
\end{table}

\fig{f1} (C)-(D) present the dependence of $T_{C}$ and $T^{\star}$ on the strontium concentration for two isotopic masses of oxygen in the LSHCO superconductor. In the whole range of concentration, we can see the completely different effect of increasing isotopic mass of oxygen on $T_{C}$ and $T^{\star}$. In the particular for the critical temperature, we have obtained the positive isotope effect (decrease in the value of $T_{C}$), and for the temperature of the pseudogap, the isotope effect is negative. 

The results collected in \tab{t1} indicate that the increase of the oxygen isotopic mass causes a slight decrease of $v$ and a slight increase of the potential $u$.

\vspace*{0.25cm}    

In the next step, we have determined a full dependence of the order parameter on temperature for the YCBCZO and LSHCO superconductors. We have considered the most interesting cases.

The results obtained for the maximum values of the critical temperature have been plotted in \fig{f2}. 

We can see that the shape of the function $\Delta\left(T\right)$ differs very significantly from the predictions of the classical BCS theory in all analyzed cases \cite{Bardeen}, \cite{Bardeen01}. 
Firstly, it should be noted that the values of the order parameter very weakly depend on temperature in the range from $0$ to $T_{C}$. 
As a result, the order parameter does not vanish for $T=T_{C}$. From the physical point of view, this fact means the existence of the pseudogap in the electronic density of states. Next, in the temperature range from $T_{C}$ to $T^{\star}$, the order parameter slightly decreases and vanishes at $T^{\star}$.

On the basis of the numerical results, \fig{f3} and \tab{t2} present the values of the zero temperature energy gap ($2\Delta\left(0\right)$) for all cases analyzed in the presented paper.

It has been found that the increase in the hole density in the YCBCZO superconductor causes a strong decrease in the value of the energy gap, and does so irrespectively of the disorder degree. It is not difficult to notice that the dependence of  $\Delta\left(0\right)$ on $p$ is clearly correlated with the shape of the function  $T^{\star}\left(p\right)$. In contrast, there is no clear relationship between the course of $\Delta\left(0\right)$ on $p$ and the form of the function $T_{C}\left(p\right)$. 

It should be noted that the results described above stand in the sharp contrast with the predictions of the classical BCS theory, in which an increase or decrease of the zero temperature energy gap is always accompanied by an increase or decrease in the critical temperature ($2\Delta\left(0\right)=3.53k_{B}T_{C}$) \cite{Carbotte}. 

\begin{figure}[ht]
\includegraphics[width=\columnwidth]{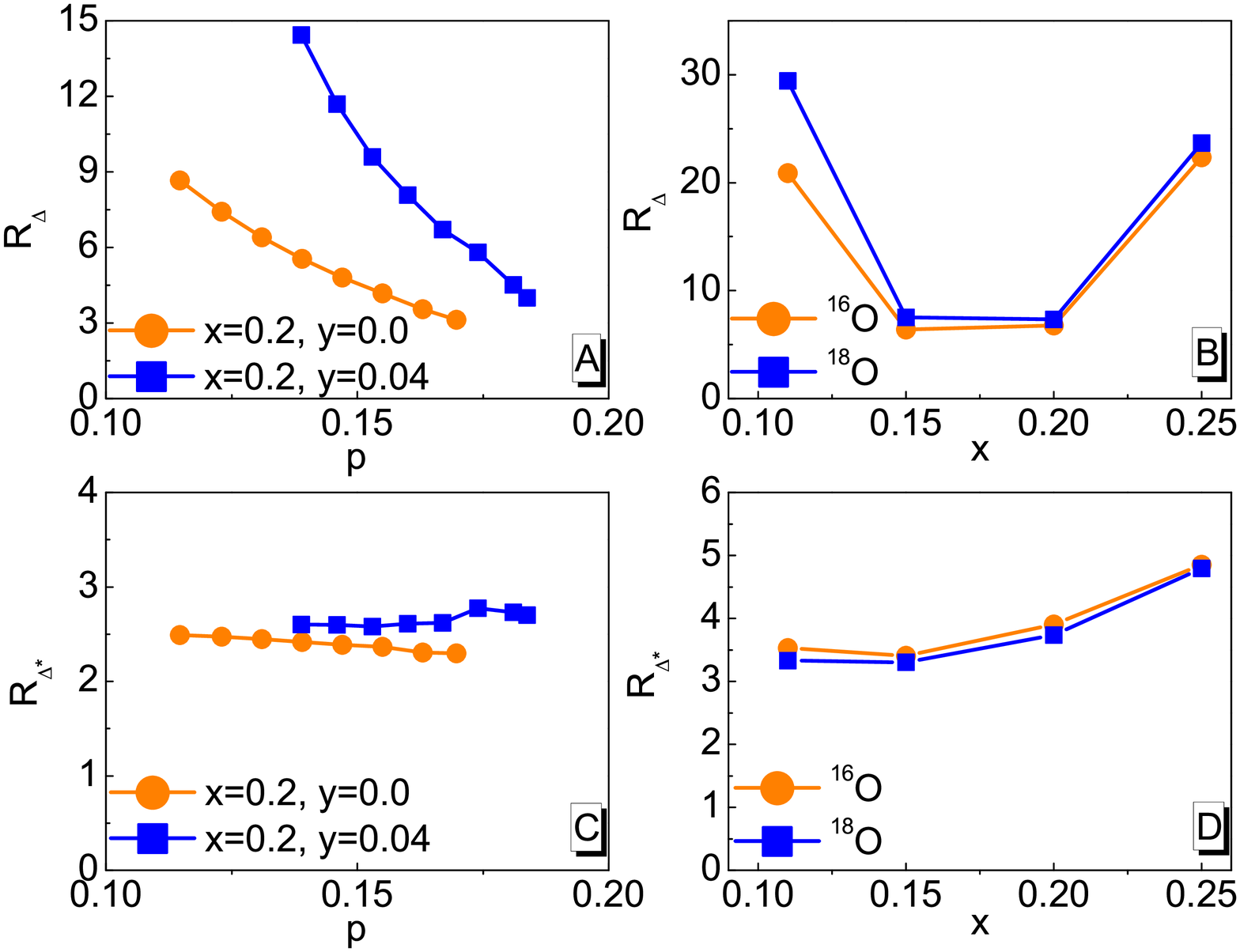}
\caption{(A)-(B) The ratio $R_{\Delta}$ for the YCBCZO and LSHCO superconductors. 
(B)-(D) The values of the ratio $R_{\Delta^{\star}}$ for the YCBCZO and LSHCO superconductors.}
\label{f4}
\end{figure}

Equally anomalous relationship exists between the increase in the in-plane disorder and the value of $2\Delta\left(0\right)$. \fig{f3} (A) clearly proves that the value of the energy gap slightly grows together with the growth of the in-plane disorder. This result is very surprising, if we take into account the fact that the value of the critical temperature drops by nearly a half at the same time (see \tab{t1}). 

The doping with strontium (the out-of-plane disorder) induces the strong decrease and then the increase in the value of the order parameter in the case of the LSHCO superconductor (\fig{f3} (B)). The obtained result comes from the analogical behaviour of the pairing potential for the renormalized electron-electron interaction with the simultaneous increase and further decrease in the potential $v$ (see \tab{t1}). Also in this case, the course of $\Delta\left(0\right)$ on x is more correlated with the shape of the function $T^{\star}\left({\rm x}\right)$ than with the function $T_{C}\left({\rm x}\right)$.  

Taking into account the influence of the oxygen isotopic mass on the value of the energy gap, it can be clearly seen that the increase in the isotopic mass 
of oxygen causes a significant increase in the value of the energy gap. The obtained result clearly correlates with the one obtained 
for the pseudogap temperature. However, it is completely inconsistent with the predictions of the BCS theory, where the isotope coefficient is positive ($\alpha=0.5$) \cite{Carbotte}.  

\vspace*{0.25cm}

The estimation of the value of the zero temperature energy gap with given $T_{C}$ and $T^{\star}$ allows in the simple way the calculation of the values of the two dimensionless parameters:

\begin{equation}
\label{r8}
R_{\Delta}\equiv \frac{2\Delta\left(0\right)}{k_{B}T_{C}}, \qquad {\rm and} \qquad R_{\Delta^{\star}}\equiv \frac{2\Delta\left(0\right)}{k_{B}T^{\star}}.
\end{equation}

The obtained results have been presented in \fig{f4} and in \tab{t2}. 

It can be noticed that the parameter $R_{\Delta}$ for both superconductors changes in the very wide range of the values, wherein the range of the values of $R_{\Delta}$ markedly broadens with the increasing in-plane disorder induced by zinc or by the isotope substitutions, where $^{16}\rm{O}$ isotope is being replaced by $^{18}\rm{O}$ isotope. It should be emphasized that the obtained result comes from the lack of the correlation between the value of the energy gap and the value of the critical temperature.  

The situation changes when we consider the parameter $R_{\Delta^{\star}}$. Based on the presented data, it is clear that the value of the energy gap varies in a similar manner as the value of the pseudogap temperature, and as such, it causes a weak dependence of the ratio $R_{\Delta^{\star}}$ on the hole density, the disorder, and the isotopic mass of oxygen. 

\vspace*{0.25cm}

In the last part of the paper, let us turn the attention toward the fact that using \eq{r7} allows the derivation of the explicit relationship between the value of the zero temperature energy gap and the pseudogap temperature. 

For this purpose, we should make use of the fact that for $T=T^{\star}$, the derivative $\frac{d\Delta}{dT}$ is unspecified. Hence, when differentiating the both sides of \eq{r7}, we get:

\begin{widetext}
\begin{equation}
\label{r9}
2u^{2}\Delta\left(I^{\left(\eta\right)}\left(\Delta,T\right)\right)^{3}\frac{d\Delta}{dT}=
-\left[v^{2}+3\left(u\Delta\right)^{2}\left(I^{\left(\eta\right)}\left(\Delta,T\right)\right)^{2}\right]
\frac{dI^{\left(\eta\right)}\left(\Delta,T\right)}{dT}.
\end{equation}

Not difficult calculations give the result:
\begin{equation}
\label{r10}
\frac{dI^{\left(\eta\right)}\left(\Delta,T\right)}{dT}=
\left[\frac{1}{k_{B}T}J_{A}^{\left(\eta\right)}\left(\Delta,T\right)-J_{B}^{\left(\eta\right)}\left(\Delta,T\right)\right]
\Delta\frac{d\Delta}{dT}-\frac{1}{4k_{B}T^{2}}J_{C}^{\left(\eta\right)}\left(\Delta,T\right),
\end{equation}
where:
\begin{equation}
\label{r11}
J_{A}^{\left(\eta\right)}\left(\Delta,T\right)\equiv
\frac{1}{N_{0}}\sum^{\omega_{0}}_{\kvec}
\frac{\eta^{4}\left({\bf k}\right)}{4\left(\varepsilon_{\kvec}^{2}+\left(\eta\left(\bf k\right)\Delta\right)^{2}\right)}
{\rm sech}^{2}\left(\frac{\beta\sqrt{\varepsilon_{\kvec}^{2}+\left(\eta\left(\bf k\right)\Delta\right)^{2}}}{2}\right),
\end{equation}
\begin{equation}
\label{r12}
J_{B}^{\left(\eta\right)}\left(\Delta,T\right)\equiv
\frac{1}{N_{0}}\sum^{\omega_{0}}_{\kvec}
\frac{\eta^{4}\left({\bf k}\right)}{2\left(\varepsilon_{\kvec}^{2}+\left(\eta\left(\bf k\right)\Delta\right)^{2}\right)^{3/2}}
\tanh\left(\frac{\beta\sqrt{\varepsilon_{\kvec}^{2}+\left(\eta\left(\bf k\right)\Delta\right)^{2}}}{2}\right),
\end{equation}
and
\begin{equation}
\label{r13}
J_{C}^{\left(\eta\right)}\left(\Delta,T\right)\equiv
\frac{1}{N_{0}}\sum^{\omega_{0}}_{\kvec}
\eta^{2}\left({\bf k}\right){\rm sech}^{2}\left(\frac{\beta\sqrt{\varepsilon_{\kvec}^{2}+\left(\eta\left(\bf k\right)\Delta\right)^{2}}}{2}\right).
\end{equation}

Substituting \eq{r10} into \eq{r9}, we have obtained the explicit expression for the derivative $\frac{d\Delta}{dT}$:
\begin{equation}
\label{r14}
\frac{d\Delta}{dT}=
\frac
{\frac{1}{4k_{B}T^{2}}\left[v^{2}+3\left(u\Delta\right)^{2}\left(I^{\left(\eta\right)}\left(\Delta,T\right)\right)^{2}\right]
J_{C}^{\left(\eta\right)}\left(\Delta,T\right)}
{
\left[
2u^{2}\Delta\left(I^{\left(\eta\right)}\left(\Delta,T\right)\right)^{3}
\left[
v^{2}+3\left(u\Delta\right)^{2}\left(I^{\left(\eta\right)}\left(\Delta,T\right)\right)^{2}
\right]
\left[
\frac{1}{k_{B}T}J_{A}^{\left(\eta\right)}\left(\Delta,T\right)-J_{B}^{\left(\eta\right)}\left(\Delta,T\right)
\right]
\right]\Delta
}.
\end{equation}
\end{widetext}

The equation that ties $\Delta\left(0\right)$ with $T^{\star}$ can be get demanding the denominator in the expression \eq{r14} to be zero. Thus,
\begin{equation}
\label{r15}
k_{B}T^{\star}=
\frac{J_{A}^{\left(\eta\right)}\left(\sigma\Delta\left(0\right),T^{\star}\right)}
{
J_{B}^{\left(\eta\right)}\left(\sigma\Delta\left(0\right),T^{\star}\right)-\frac{2u^{2}
\left[I^{\left(\eta\right)}\left(\sigma\Delta\left(0\right),T^{\star}\right)\right]^{3}}
{v^{2}+3u^{2}\left[\sigma\Delta\left(0\right)\right]^{2}\left[I^{\left(\eta\right)}\left(\sigma\Delta\left(0\right),T^{\star}\right)\right]^{2}}
},
\end{equation}
where the parameter $\sigma$ equals approximately $0.76$. 

Let us note that the value of $\sigma$ has been chosen in such a way that it can, in the most precise way, allow the reproduction of the numerical results (see \tab{t3}).

\begin{table}[t]
\renewcommand*{\tablename}{\small{Tab.}}
\renewcommand*{\baselinestretch}{1}
\caption{\label{t3}The results of the calculations for the values of the parameter $\sigma=0.76$. 
The symbol $\Delta T^{\star}$ has been defined with the help of the formula:
$\Delta T^{\star}\equiv 100\%*\left(\left[T^{\star}\right]_{{\rm n}}-\left[T^{\star}\right]_{{\rm Eq.}}\right)/\left[T^{\star}\right]_{{\rm n}}$, where $\left[T^{\star}\right]_{{\rm n}}$ and $\left[T^{\star}\right]_{{\rm Eq.}}$ denote the values of the pseudogap temperature obtained numerically or by using Eq. (15)  respectively.}
\begin{center}
\begin{tabular}{|c|c|c|c|c|}\hline
                          &           &        &        &     \\
{\bf Material} & 
${\bf Type}$   & 
${\bf \left[T^{\star}\right]_{n}}$ ${\bf \left[K\right]}$&
${\bf \left[T^{\star}\right]_{Eq.}}$ ${\bf \left[K\right]}$&
$|\Delta T^{\star}|$ $\%$ \\
                            &             &            &          &\\
\hline
                            &             &            &          & \\
YCBCZO    \\ (x=0.2, y=0)   & $p=0.115$   &   243.9    &   279.0  & 14.4  \\
                            & $p=0.123$   &   224.4    &   258.1  & 15.0  \\
                            & $p=0.131$   &   205.5    &   233.3  & 13.5  \\
                            & $p=0.139$   &   186.7    &   211.6  & 13.3  \\
                            & $p=0.147$   &   167.8    &   191.3  & 14.0  \\
                            & $p=0.155$   &   148.9    &   169.0  & 13.5  \\
                            & $p=0.163$   &   130.0    &   147.8  & 13.7  \\
                            & $p=0.170$   &   114.2    &   136.4  & 19.4  \\
                            &             &            &          &       \\
\hline
                            &             &            &          &        \\
YCBCZO    \\ (x=0.2, y=0.04)& $p=0.139$   &   196.5    &   214.5  &  9.2   \\ 
                            & $p=0.146$   &   175.2    &   189.2  &  8.0   \\
                            & $p=0.153$   &   154.4    &   163.8  &  6.1   \\
                            & $p=0.160$   &   133.5    &   140.7  &  5.4   \\
                            & $p=0.167$   &   112.7    &   117.5  &  4.3   \\
                            & $p=0.174$   &    91.9    &   101.1  &  10.1  \\
                            & $p=0.181$   &    71.1    &    77.1  &  8.4   \\
                            & $p=0.184$   &    62.9    &    68.3  &  8.6   \\
                            &        &        &         &      \\

\hline
                            &        &        &         &      \\
LSHCO    \\ ($^{16}\rm{O}$) & x=0.11 &  82.2  &  69.1   &  15.9\\
                            & x=0.15 &  61.0  &  64.6   &   5.9\\
                            & x=0.20 &  49.2  &  54.0   &   9.8\\
                            & x=0.25 &  46.5  &   -     &   -  \\
                            &        &        &         &      \\
\hline
                            &        &         &         &    \\
LSHCO    \\ ($^{18}\rm{O}$) & x=0.11 &  103.5  &  95.9   & 7.3\\
                            & x=0.15 &   70.0  &  69.9   & 0.1\\
                            & x=0.20 &   53.7  &  56.6   & 5.4\\
                            & x=0.25 &   48.4  &   -     &  - \\
                            &        &         &         &    \\
\hline
\end{tabular}
\end{center}
\end{table}

Additionally, it can be seen that in the boundary of $v/u\rightarrow 0$, the equation (15) takes the particularly simple form:
\begin{equation}
\label{r16}
k_{B}T^{\star}=
\frac{J_{A}^{\left(\eta\right)}\left(\sigma\Delta\left(0\right),T^{\star}\right)}
{
J_{B}^{\left(\eta\right)}\left(\sigma\Delta\left(0\right),T^{\star}\right)
-\frac{2}{3}\frac{1}{\left[\sigma\Delta\left(0\right)\right]^{2}}
I^{\left(\eta\right)}\left(\sigma\Delta\left(0\right),T^{\star}\right)
}.
\end{equation}

In conclusion: the study has examined the effect of the hole density, the out-of-plane or in-plane disorder, and the isotopic mass of oxygen on the value of the zero temperature energy gap in the YCBCZO and LSHCO superconductors. 

It has been found that - regardless of the type of the studied material - the zero temperature energy gap is closely correlated with the pseudogap temperature. In contrast, there was no correlation between $2\Delta\left(0\right)$ and the critical temperature. 

The obtained results indicate that the value of the ratio $R_{\Delta}$ can vary and can widely depend on the deviations from the initial chemical composition. On the other hand, changes in the value of the parameter $R_{\Delta^{\star}}$ will be rather small. 

In the presented paper, we have explicitly included all important numerical results, and for that reason we strongly encourage all readers verify them quantitatively by means of the available experimental methods.

\vspace*{0.25 cm}
\begin{acknowledgments}

The authors would like to thank Prof. K. Dzili{\'n}ski for creating excellent working conditions.

R.S. would like to express his gratitude to Sisi and Okta for the exhaustive scientific pieces of advice related 
to the topic tackled by the presented work.

Additionally, the authors are grateful to the Cz{\c{e}}stochowa University of Technology - MSK CzestMAN for granting access 
to the computing infrastructure built in the project No. POIG.02.03.00-00-028/08 "PLATON - Science Services Platform".

\end{acknowledgments}


%
\end{document}